\documentclass[pra,floatfix,showpacs,twocolumn]{revtex4}
\usepackage{amstext}
\usepackage{amsmath}
\usepackage{amssymb}
\usepackage{graphicx}
\usepackage{latexsym}
\usepackage{bm}
\providecommand{\U}[1]{\protect\rule{.1in}{.1in}}
\begin{document}

\title{Dynamic method to distinguish between left- and right-handed chiral
molecules}
\author{Yong Li}
\author{C. Bruder}
\affiliation{Department of Physics, University of Basel, Klingelbergstrasse
82, 4056-Basel, Switzerland}
\date{\today }

\begin{abstract}
We study quantum systems with broken symmetry that can be modelled as
cyclic three-level atoms with coexisting one- and two-photon
transitions. They can be selectively optically excited to any
state. As an example, we show that left- and right-handed chiral
molecules starting in the same initial states can evolve into
different final states by a purely dynamic transfer process. That
means, left- and right-handed molecules can be distinguished purely
dynamically.
\end{abstract}

\pacs{
33.15.Bh, 
33.80.-b, 
42.50.Hz, 
}

\maketitle

\section{Introduction}

In quantum systems with well-defined parities,
one- and two-photon processes cannot coexist due to the
electric-dipole selection rule \cite{group}. However, this does not
apply for quantum systems with a broken symmetry, e.g., chiral
molecules whose symmetry is broken naturally, asymmetric quantum
wells, or artificial ``atoms'' with broken symmetry, in which optical
transitions will be allowed between any two levels. Recently,
symmetry-broken artificial atoms \cite{Liu05} in a superconducting
flux qubit circuit were investigated. By biasing the external magnetic
flux $\Phi_{e}$ away from $ \Phi _{0}/2$ ($\Phi _{0}=h/2e$ is the
magnetic flux quantum, $h$ is the Planck constant and $e$ the
electronic charge), the symmetries of both the potential and the
interaction Hamiltonian can be broken. In this case, when only the
lowest three levels are considered, the flux qubit circuit is a
cyclic-type (or $\Delta$-type) system, which is different from the
usual $\Xi$-type (or ladder-type) system in which the optical
dipole-selection rule does not allow a transition between the lowest
and the uppermost of the three levels. By investigating the
generalized stimulated Raman adiabatic passage (STIRAP \cite{STIRAP}),
one can achieve a pulse-phase-sensitive adiabatic manipulation of
quantum states. The population can be cyclically transferred by
controlling the amplitudes and/or phases of the coupling pulses. A
cyclic system involving the coupling of nonclassical (quantized)
optical fields has been investigated in \cite{nonclassic}.
Adiabatical (or dynamical) quantum information transfer between the
collective excitations of a cyclic ``atomic'' ensemble and quantized
optical fields has also been studied \cite{collective}.

The existence of chiral molecules is one of the fundamental broken
symmetries in nature. Chiral molecules can be modelled as being in a
mirror-symmetric double-well potential \cite{Hund:1927,Harris:1978}:
left-handed states are in the left well and right-handed ones in the
right. Strictly speaking, left-/right- handed states in the left/right wells
are not the eigenstates of double-well potential due to the tunnelling
between the left and right wells. But the tunnelling rates are usually
very small, so left-/right handed states can be considered as stable
eigenstates of chiral-molecule systems. Chiral purification and
discrimination
\cite{Bodenhoefer:1997,McKendry:1998,Rikken:2000,Zepik:2002,Bielski}
of a mixture of chiral molecules are among the most important and
difficult tasks in chemistry. The usual purification methods for
chiral molecules are based on asymmetric synthesis, e.g., gas or
liquid chromatographic methods \cite{asymmetric-synthesis}. Methods of
achieving purification solely by optical means have also been
investigated theoretically
\cite{Shapiro,Salam:1998,Fujimura:1999}. 
Recently, Kr\'{a}l \textit{et al.} \cite{Kral01,Kral03} proposed an 
optical cyclic population
transfer scheme to distinguish mixed left- and right-handed
molecules. Each kind of chiral molecule (left-handed or right-handed)
can be modelled as a cyclic system if only the three lowest levels are
considered. The left-handed cyclic system behaves like the
right-handed one except the intrinsically-different total phases of
the three Rabi frequencies if three transitions are coupled to three
optical fields, respectively. The idea was to optically transfer
chiral molecules which are in their initial respective ground states
to final levels at different energies. This population transfer was
achieved by adiabatical (or diabatical) optically coupled processes.

In another development, there has been considerable interest to
implement pseudo-spin dependent optically-induced gauge potentials for
cold atoms to achieve a spatial separation of pseudo-spin states
\cite{zhu,liu}. Based on this scheme, we
recently considered consequences of the induced gauge potential in
systems of cold chiral molecules that manifest themselves as a
generalized Stern-Gerlach effect \cite{SG-effect}, 
where the orbital motion of mixed chiral molecules will be
chirality-dependent and pseudo-spin-dependent. Thus, it can be used to
distinguish molecules with different chiralities, suggesting a
discrimination method to separate chiral mixtures.

In this paper, we will use a dynamical (rotation) method to transfer
the states of cyclic three-level systems with broken symmetry by
applying ultrashort $\pi$- or $\pm \pi/2$ optical
pulses. Interestingly, we can distinguish mixed left- and right-handed
molecules (in their respective ground states) by dynamically driving
them to different-energy final states.

\section{Population Transfer via dynamic process}

The general cyclic three-level system coupled with three classical
optical fields respectively can be described by the
Hamiltonian \cite{Kral01} in the rotating-wave approximation ($\hbar =1$)
\begin{equation}
H_{ori}=\sum_{j=1}^{3}\omega _{j}\left\vert j\right\rangle
+\sum_{i>j=1}^{3}\left( \Omega _{ij}(t)e^{-i\omega _{ij}t}\left\vert
i\right\rangle \left\langle j\right\vert +\text{H.c.}\right) ,  \label{hamil}
\end{equation}
where $\omega _{j}$ are the energies of the states
$\vert j\rangle$, $\omega _{ij}$ are the frequencies of the optical
fields coupling to the transition $\left\vert i\right\rangle
\leftrightarrow \left\vert j\right\rangle$, and
$\Omega_{ij}(t)=\vec \mu_{ij}\cdot \vec E_{ij}$ are
the Rabi frequencies which can be controlled by varying the field
strength $\vec E_{ij}$; $\vec \mu_{ij}$ is
the dipole matrix element between the states $|i\rangle$ and
$|j\rangle$. The analytic instantaneous eigenfrequencies and
eigenfunctions have been evaluated in Refs. \cite{Kral01,Liu05}. At
resonance, $\omega_{ij}\equiv \omega_{i}-\omega_{j}$, the above
Hamiltonian can be rewritten in the interaction picture as
\begin{equation}
H=\sum_{i>j=1}^{3}\Omega _{ij}(t)\left\vert i\right\rangle \left\langle
j\right\vert +\text{H.c.}.  \label{hamil-res}
\end{equation}

\begin{figure}[ht]
\includegraphics[height=4 cm,width=7 cm]{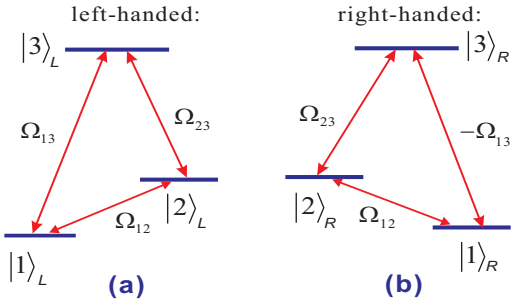}
\caption{(Color online) Model of left-(a) and right-(b) handed chiral
molecule by a three-level $\Delta$-type (or cyclic) system, which is
resonantly coupled to three classic optical fields with Rabi
frequencies $\Omega_{12}$, $\Omega_{23}$, and $\pm\Omega_{13}$.}
\label{a1}
\end{figure}

The goal of this paper is to present a purely dynamic method to
separate left- and right-handed chiral molecules. 
If left- and right-handed
three-level cyclic molecules are driven by three optical fields,
one or all of the three Rabi frequencies $\Omega_{ij}(t)$ for the
two kinds of chiral molecules differ by a sign \cite
{Kral01,Kral03}. We will specify our model by choosing the
coupling strengths (that is, Rabi frequencies) of the
left-/right- handed molecules as $\Omega _{ij}^{L}(t)\equiv \Omega
_{ij}(t)$, and $\Omega _{13}^{R}(t)\equiv -\Omega _{13}(t)$,
$\Omega _{12}^{R}(t)\equiv \Omega _{12}(t)$,$\ \Omega
_{23}^{R}(t)\equiv \Omega _{23}(t)$ (see Fig.~\ref{a1}). Here, the
superscripts $L$ [$R$] refer to left-handed [right-handed]
molecules.

\begin{figure}[tbp]
\includegraphics[height=4 cm,width=7 cm]{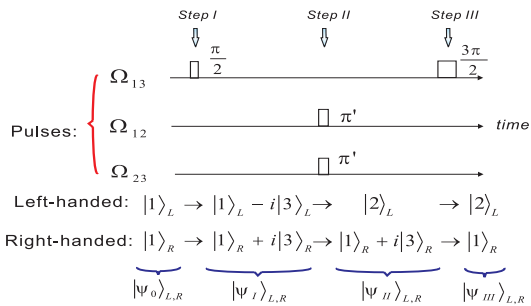}
\caption{(Color online) Schematic representation of the three
steps to discriminate two kinds of chiral molecules. All the
operations consist of simple $\pm \pi/2$ and $\pi $ rotation
pulses of three optical coupling strengths (Rabi frequencies).
Here, the notation $\pi^{\prime}$ is used to indicate a $\pi$
rotation pulse for the effective Rabi frequency
$\Omega_{\text{eff}}(t)$ ($=\protect\sqrt{2}|\Omega _{12}(t)|=
\protect\sqrt{2}|\Omega _{23}(t)|=\protect\sqrt{2}\Omega
_{0}(t)$), and {\it not} for the Rabi frequencies $\Omega
_{12}(t)$ and $\Omega _{23}(t)$. } \label{a2}
\end{figure}

We now consider an ensemble of left- and right-handed molecules
prepared in their ground states
$\vert 1\rangle_{L}$ and $\vert1\rangle _{R}$ respectively.
We aim to distinguish the two kinds of molecules by transferring them
to different-energy final states. Our separation protocol consists of
the following steps (see Fig.~\ref{a2}):

\emph{Step I}: At time $t=0$, we start with a
pump pulse $\Omega _{13}(t)$
for the left-handed molecules. The corresponding Hamiltonian will reduce to
\begin{equation}
H_L^I=\Omega _{13}(t)\left\vert 1\right\rangle _{LL}\left\langle
3\right\vert +\text{H.c.}.  \label{hamil-left1}
\end{equation}
Here we control the pulse to cause a $\pi/2$ rotation \cite{rotation}
for the left-handed molecules,
\begin{equation}
\left\vert \psi _{0}\right\rangle _{L}\equiv \left\vert 1\right\rangle
_{L}\rightarrow \left\vert \psi _{I}\right\rangle _{L}\equiv
\frac{1}{\sqrt{2}}\left( \left\vert 1\right\rangle _{L}- i\left\vert
3\right\rangle_{L}\right) .
\end{equation}
For the right-handed ones, the Rabi frequency is $-\Omega_{13}(t)$, which
leads to a $-\pi/2$ rotation
\begin{equation}
\left\vert \psi _{0}\right\rangle _{R}\equiv \left\vert 1\right\rangle
_{R}\rightarrow \left\vert \psi _{I}\right\rangle _{R}\equiv \frac{1}
{\sqrt{2}}\left( \left\vert 1\right\rangle _{R}+i\left\vert 3\right\rangle
_{R}\right)
\end{equation}
with the corresponding Hamiltonian
\begin{equation}
H_R^I=-\Omega _{13}(t)\left\vert 1\right\rangle _{RR}\left\langle
3\right\vert +\text{H.c.}.  \label{hamil-right1}
\end{equation}

\emph{Step II}: Now we keep $\Omega _{13}(t)\equiv 0$ and add two pump
pulses of $\Omega _{12}(t)$ and $\Omega _{23}(t)$, which satisfy
\begin{equation}
\Omega _{12}(t) =i|\Omega _{12}(t)|\equiv i\Omega _{0}(t)=i\Omega _{23}(t).
\end{equation}
Hence, the Hamiltonian (\ref{hamil-res}) for the left-handed molecules reads
\begin{equation}
H_{L}^{II}=\Omega _{\text{eff}}(t)\left( \left\vert B\right\rangle _{LL}\left\langle
2\right\vert +\text{H.c.}\right) ,  \label{hamil-d2}
\end{equation}
where $\left\vert B\right\rangle _{L}=(i\left\vert
1\right\rangle _{L}+\left\vert 3\right\rangle _{L})/\sqrt{2}$, and
$\Omega_{\text{eff}}(t)=\sqrt{2}\Omega_{0}(t)$.
The instantaneous eigenfunctions for Eq.~(\ref{hamil-d2}) are
\begin{eqnarray}
\left\vert E_{0}\right\rangle _{L} &=&\frac{1}{\sqrt{2}}\left( \left\vert
1\right\rangle _{L}+i\left\vert 3\right\rangle _{L}\right) ,  \notag \\
\text{ }\left\vert E_{\pm }\right\rangle _{L} &=&\frac{1}{\sqrt{2}}\left(
\left\vert 2\right\rangle _{L}\pm \left\vert B\right\rangle _{L}\right)
\label{eigen-state}
\end{eqnarray}
with the corresponding eigenvalues
\begin{equation}
E_{0}=0,\quad E_{\pm }=\pm \Omega_{\text{eff}}(t).
\label{values}
\end{equation}

Hence, for the left-handed molecule, the corresponding time-evolution of the state is given as
\begin{eqnarray}
\left\vert \psi_{II}\right\rangle_{L}&=& e^{-i\int_{0}^{t}H_{L}^{II}(t^{\prime })dt^{\prime}}\left\vert \psi _{I}\right\rangle _{L} 
\notag \\
&=&\frac{-i}{\sqrt{2}}\left( e^{-i\eta (t)}\left\vert E_{+}\right\rangle
_{L}-e^{i\eta (t)}\left\vert E_{-}\right\rangle _{L}\right) ,
\end{eqnarray}
where
$\eta (t):=\int_{0}^{t}\Omega _{\text{eff}}(t^{\prime })dt^{\prime}$.
If $\Omega _{\text{eff}}(t)$ is chosen to cause a rotation by $\pi$
corresponding
to the transition $\vert 2\rangle _{L}\leftrightarrow \vert B\rangle_{L}$,
that is, $\eta(t)=\pi/2$, the evolved state is
\begin{equation}
\vert \psi _{II}\rangle _{L}=-\left\vert 2\right\rangle _{L}.
\end{equation}

The same treatment applies to the right-handed molecules by replacing
``L'' by ``R'' in Eqs.~(\ref{hamil-d2}-\ref{eigen-state}). Since the
state
$\vert \psi _{I}\rangle _{R} $ $=( \vert 1\rangle_{R}+i\vert 3\rangle_{R})
/\sqrt{2}$ is always the dark
state for the Hamiltonian $H_{R}^{II}=\Omega _{\text{eff}}(t)\left( \left\vert
B\right\rangle _{RR}\left\langle 2\right\vert +\text{H.c.}\right)$,
it is invariant during the $\pi$ rotation in this step. Thus, we obtain
the following state after step II:
\begin{equation}
\left\vert \psi _{II}\right\rangle _{R}\equiv \left\vert \psi
_{I}\right\rangle _{R}=\frac{1}{\sqrt{2}}\left( \left\vert 1\right\rangle
_{R}+i\left\vert 3\right\rangle _{R}\right)
\end{equation}
for the right-handed molecules.

\emph{Step III}: Now we keep $\Omega _{12}(t)=\Omega _{23}(t)\equiv 0$
and apply the pump pulse $\Omega _{13}(t)$ leading to a $3\pi /2$
rotation for the left-handed molecules (the corresponding Rabi
frequency is $-\Omega _{13}(t)$ with a $-3\pi /2$ (or $\pi /2$)
rotation for the right-handed ones). The left-handed molecular
state $\vert 2\rangle _{L}$ remains unchanged:
\begin{equation}
\left\vert \psi _{III}\right\rangle _{L}=\left\vert \psi _{II}\right\rangle
_{L}=-\left\vert 2\right\rangle _{L}.
\end{equation}
On the other hand, the $-3\pi /2$ rotation pulse acting on the
right-handed molecules will transfer the state
$\vert \psi_{II}\rangle_{R}$ to the final state
\begin{equation}
\left\vert \psi _{III}\right\rangle _{R}=\left\vert 1\right\rangle _{R}.
\end{equation}
Here, the corresponding Hamiltonians for the chiral molecules are
again given by Eqs. (\ref{hamil-left1},\ref{hamil-right1}),
respectively.

The above protocol, which includes several simple ultrashort rotations,
will bring the initial state $\vert 1\rangle _{L}\rightarrow
\vert 2\rangle _{L}$ for the left-handed molecules and
$\vert 1\rangle _{R}\rightarrow \vert 1\rangle _{R}$
for the right-handed ones. The fact that differently-handed chiral molecules
are transferred to different long-lived states means perfect discrimination
of both kinds of chiral molecules in our idealized treatment.

\begin{figure}[tbp]
\centerline{\includegraphics[width=7.5cm,height=4cm]{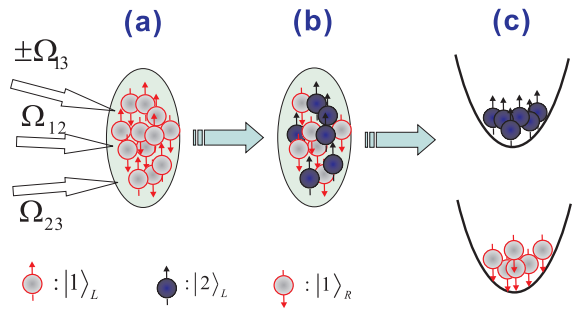}}
\caption{(Color online) Illustration of the separation procedure.
(a) Initially, all the (oriented) chiral molecules are prepared in
the ground states $\vert 1\rangle$ before the coupling to
the optical fields (here the optical fields can be coplanar or not). 
The labels of spin-up [down] refer to the left-
[right-] handed molecules. (b) After driving the optical pulses in
the protocol as shown in Fig.~\protect\ref{a2}, the chiral
molecules are dynamically transferred to the final state $\vert
2\rangle_{L}$ and $\vert 1\rangle_{R}$, respectively. The dark
[bright] spheres indicate the states $\vert 2\rangle$ [$\vert
1\rangle $]. (c) The different-handed chiral molecules may now be
separated, e.g., by ionization or choosing different trapping
potentials for different energy levels.} \label{a3}
\end{figure}

Figure \ref{a3} shows an illustration of this separation procedure.
Initially, all the molecules are assumed to be in the ground state
$\vert 1\rangle_{L,R}$ as shown in Fig.~\ref{a3}(a). By using simple
optical pulses, i.e., switching on/off the optical fields as shown in
Fig.~\ref{a2}, the two kinds of chiral molecules are dynamically
transferred from the initial state $\vert 1\rangle_{L,R}$ to the final
state $\vert 2\rangle _{L}$ and $\vert 1\rangle _{R}$, respectively
(see Fig.~\ref{a3}(b)). These different states may be separated (see
Fig.~\ref{a3}(c)) using a variety of energy-dependent processes, such
as ionization, followed by extraction of the ions by an electric field as
suggested in Ref.~\cite{Kral01}.

The scheme proposed here is different from the generalized
Stern-Gerlach effect \cite{SG-effect} in the following respects: 
(i) There is no need to invoke the spatial motion of the molecules in the
scheme presented here; it is enough to consider their internal
dynamics. 
(ii) The internal degree of freedom of a three-level system is reduced
to the lower two (pseudo-spin) states in \cite{SG-effect} by assuming
a large detuning. Here, we discuss the molecular dynamics in the space
of three levels with all the optical coupling resonant to the
corresponding transition.
(iii) Different chiral molecules can be separated spatially; here, we
only focus on how to distinguish different chiral molecules according
to the different final states as in Ref.~\cite{Kral01,Kral03}.

The present protocol is also different from the methods described in
Refs.~\cite{Kral01,Kral03}, where molecules with different chirality
are distinguished by adiabatical processes via controlling three
overlapping optical fields. The dynamical rotation method proposed
here provides another possibility to distinguish mixed chiral
molecules by using several simple optical ultrashort pulses.
In contrast to the scheme described in Ref.~\cite{Kral03}, we assume
the molecules to be pre-oriented (similar to
Refs.~\cite{Kral01,SG-effect}) in order to ensure that the Rabi
frequencies are identical for all the molecules in the mixture (up to
a phase factor) when coupling to a same optical field.

In the present separation scheme as well as similar ones
\cite{Kral01,Kral03,SG-effect}, the molecular temperature is assumed
to be low enough such that the molecules are prepared in the ground state
initially.  We neglect finite-temperature effects and molecular
interactions.

In the protocol described above, all three kinds of optical fields
couple resonantly to the molecular transitions. That means the optical
fields satisfy the energy conservation condition
$\omega_{12}+\omega_{23}=\omega_{13}$. This resembles sum-frequency
generation in nonlinear media \cite{SFG}, which requires energy
conservation as well as momentum conservation (along a certain
direction). Sum-frequency generation is one of the techniques used to
probe (but not separate) molecular chirality \cite{SFG-chiral}. The
separation protocol described in our paper is not related to
sum-frequency generation and is not constrained by the momentum
conservation of optical wave-vectors (that is why the three optical
fields can be coplanar or not).

Recently, many studies about preparing, measuring and teleporting
superpositions of chiral states \cite{superposition} have appeared in
the literature. In these works, superposition states are obtained by
involving a symmetric or antisymmetric higher excited state which can
couple to the left-handed or right-handed states by optical fields. In
this paper, we aim to separate mixtures of left-handed and
right-handed chiral molecules. So the external excited state is not
introduced, and superpositions of chiral states are not considered
here.

Decoherence processes are the biggest obstacle for any quantum state
transfer operation. Due to the fact that our protocol only uses
dynamic ultrashort-pulse operations and does not require adiabaticity
assumptions, our method will work faster and make decoherence effects
less important.

\section{Conclusion}
In conclusion, we have presented a protocol to separate left- and
right-handed molecules. This protocol, which consists of several
simple dynamical rotation operations of optical pulses, can be
implemented on a short time scale compared with the decoherence time
due to spontaneous emission. We would like to remark that
all three kinds of optical couplings have to be used (though
not at the same time) during the rotation operations in this
separation protocol. Physically, this is because the two kinds of
chiral molecules can be distinguished only by the total phase of the
three Rabi frequencies (which differ by $\pi$). If only two kinds of
optical couplings are used, the molecular system will reduce to the
$V$-, $\Lambda$- or ladder-type three-level system, which cannot be
used to distinguish different chiralities.

This work was supported by the Swiss NSF and the NCCR Nanoscience.
Y.L. also acknowledges support from NSFC through Grant No. 10574133.


\begin{thebibliography}{19}
\bibitem{rotation} Here, the $\pi/2$ rotation means
$\int_{0}^{t}\Omega_{13}(t^{\prime})dt^{\prime }=\pi/4$,
which can be accomplished in an ultrashort time. The simplest case is to
keep $\Omega _{13}(t)\equiv \Omega _{c}$
constant for a time $\pi/(4\Omega _{c})$, which is much shorter
than the decoherence time to make relaxation effects negligible.
\end{thebibliography}
\end{document}